\documentclass[twocolumn,prl]{revtex4}
\usepackage{epsfig,amssymb,amsmath,graphicx}
\begin{document}
\title{The c-axis charge traveling wave in coupled system of Josephson junctions}

\author{ Yu.M.Shukrinov~$^{1}$}
\author{M.Hamdipour~$^{1,2}$ }
\address{$^{1}$ BLTP, JINR, Dubna, Moscow Region, 141980, Russia\\
$^{2}$Institute for Advanced Studies in Basic Sciences, P.O.Box 45195-1159, Zanjan, Iran}

\begin{abstract}
We demonstrate a manifestation of the charge traveling wave  along the $c$-axis (TW) in current voltage
characteristics  of coupled Josephson junctions  in high-$T_c$ superconductors. The branches related to the TW
with different wavelengths are found for the stacks with different number of Josephson junctions at different
values of system's parameters. Transitions between the TW branches  and  the outermost  branch are observed.
Time dependence of the electric charge in the superconducting layers and charge-charge correlation functions for
TW and  outermost branches  show different behavior with bias current. We propose an experimental testing of the
TW by microwave irradiation.
\end{abstract} \maketitle

Many physical and biological objects may be considered as  coupled nonlinear oscillators.\cite{strogatz93}
Particularly, intrinsic Josephson junctions (JJ) in HTSC are studied from this point of view
often.\cite{kleiner} An attractive feature of  intrinsic JJ is a  generation of coherent radiation in terahertz
region and it is  investigated intensively.\cite{kleiner-science07,tsujimoto10} A coupling between junctions
plays an important role. It leads to multiple branch structure in the current voltage characteristics (CVC) with
additional to single junction parameters like the breakpoint  (BP) current, the BP region width  and the
transition currents between branches.\cite{sm-sust07,matsumoto99,sm-prl07,smp-prb07} It was shown  that the BP
in CVC is a manifestation of the parametric resonance in this system when the Josephson oscillations create a
standing longitudinal plasma wave perpendicular to the superconducting layers. The parametric resonance features
were observed experimentally recently.\cite{iso-apl08}

A well known feature of single long Josephson junction is the traveling wave along the barrier.\cite{kleiner} It
was noted\cite{koyama96,machida98} that the system of capacitively coupled JJ has a plane wave which corresponds
to the longitudinal Josephson plasma propagating perpendicular to the junctions. The sharp resonance peak in the
microwave absorption spectrum obtained in Bi-2212 by Matsuda et al.\cite{matsuda95} has been identified with
this collective Josephson plasma mode. Instability of the longitudinal plasma oscillations was discussed in
Ref.\cite{machida98} and it was proposed that the multiple branches are caused by the freezing of the collective
LPW into the static charge density wave.

In this letter we demonstrate  the additional branch in CVC which corresponds to the charge traveling wave along
c-axis (TW) when the junctions are in rotating states (phase differences are not small). Such branches were not
investigated before. Transitions in hysteresis region from the outermost branch (OB) to the traveling wave
branch  (TWB) and between the TWB with different wavelengths  are found. We demonstrate the appearance of the
TWB at $I=I_c$ as well which originates  from different branches of CVC including the zero voltage branch. A
different dynamics of correlation functions for the TWB and for the OB is shown. We find the effect of the
microwave radiation on TWB, which allows the experimental testing of this branch.

To simulate the CVC we investigate the phase dynamics  in the framework of the capacitively coupled Josephson
junctions with diffusion current (CCJJ+DC) model,\cite{machida00,sms-physC06} which is determined by the system
of dynamical equations$\frac{d}{dt}V_{l}=I-\sin \varphi_{l} -\beta\frac{d\varphi_{l}}{dt},
\frac{d}{dt}\varphi_{l}= V_l - \alpha (V_{l+1}+ V_{l-1} - 2V_{l})$ for the gauge-invariant phase differences
$\varphi_l(t)$  between superconducting layers ($S$-layers). Time $t$ is normalized to the inverse  plasma
frequency $\omega^{-1}_p$ ($\omega_p^2=2eI_c/\hbar C$), the voltage - to the value $V_0=\hbar \omega_p/2e$, the
current -  to the critical current $I=I_c$.  We consider the periodic boundary conditions in this paper. The
structure of the simulated CVC in the CCJJ+DC model  is equidistant in agreement with the experimental
results.\cite{schlenga98} The details of simulation procedure are presented in Refs\cite{smp-prb07, shk-prb09}

\begin{figure}[ht]
 \centering
\includegraphics[height=60mm]{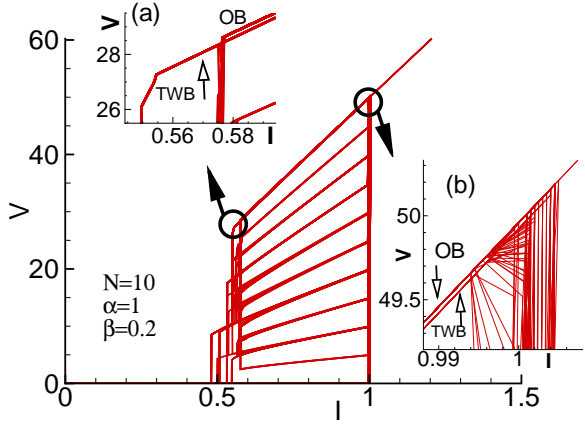}
\caption{(Color online) Manifestation of the TWB in CVC of the stacks with 10 JJ. The insets  enlarges the parts
of CVC: (a) near transition from the OB to another branch; (b) near transition at $I=I_c$, where the TWB is
created.}
 \label{1}
\end{figure}

The manifestation of the TWB is shown in Fig.~\ref{1}, where we present  the total branch structure of CVC for
the stack with 10 coupled JJ at $\alpha=1$, $\beta=0.2$. It was obtained by multiple sweeping of the bias
current through the stack.\cite{sm-physC06,sm-sust07,smp-prb07} This CVC is characterized by switching all
junctions to the rotating (R-) state at $I=I_c$, large hysteresis, and multiple branching in the hysteresis
region by switching the JJ to the oscillating (O-) state with the decrease of the bias current.

The inset (a) enlarges the part of CVC corresponding to the transition from the OB to the another branch. As we
can see, there is one more branch additionally to the  OB. We prove below that this is TWB. Such TWB was
manifested in CVC of the stack with 10 JJ in Ref.\cite{sm-sust07} (see Fig.5), but it was not mentioned at that
time.

To see the origin of  TWB we investigate a transition region in CVC near $I_c$. In during of the sweeping bias
current procedure we  observe the jumps to the OB and TWB both from the zero voltage state directly  and from
the other branches of CVC. The inset (b) enlarges the part of CVC corresponding to this transition region.

\begin{figure}
\includegraphics[width=35mm]{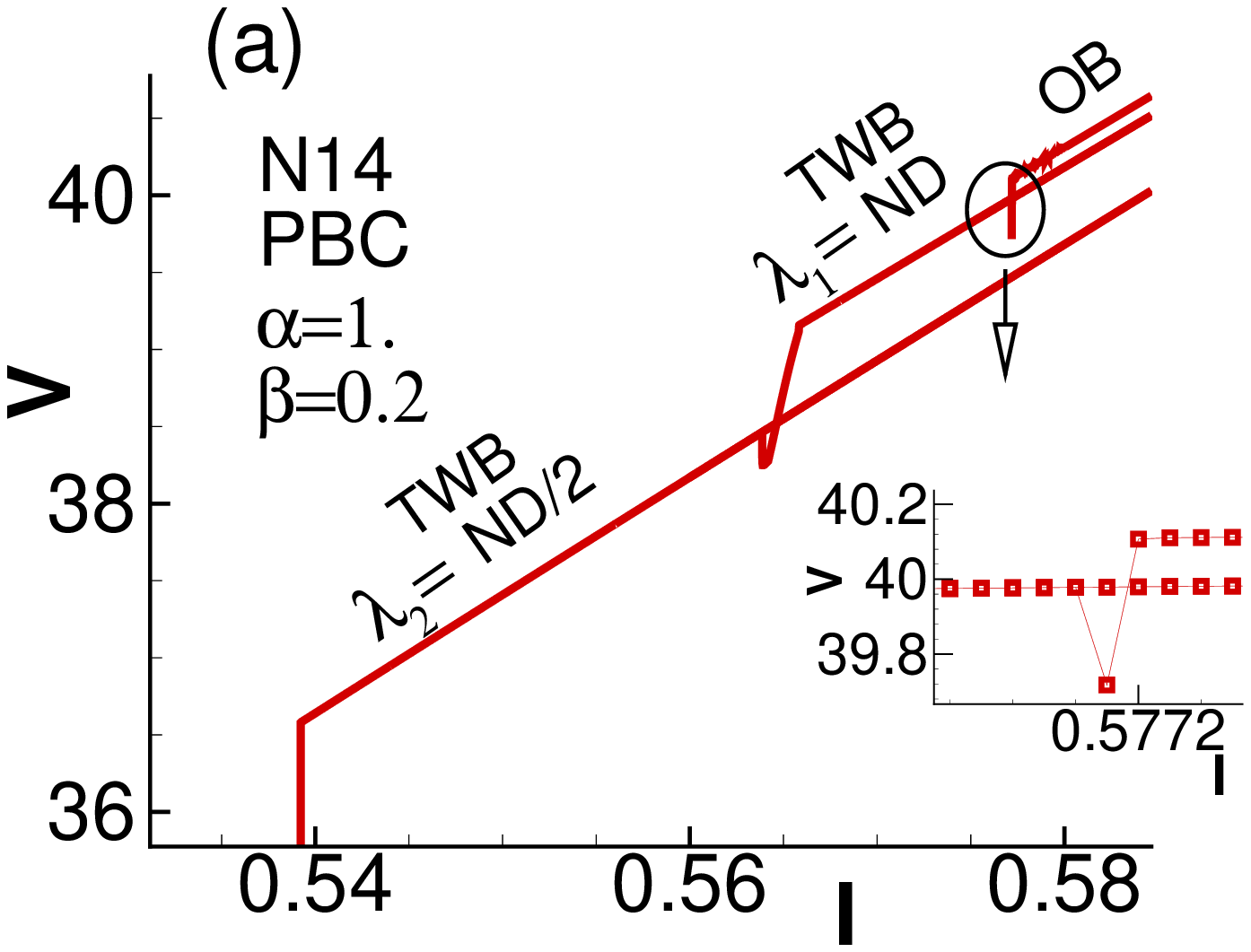}
\includegraphics[width=35mm]{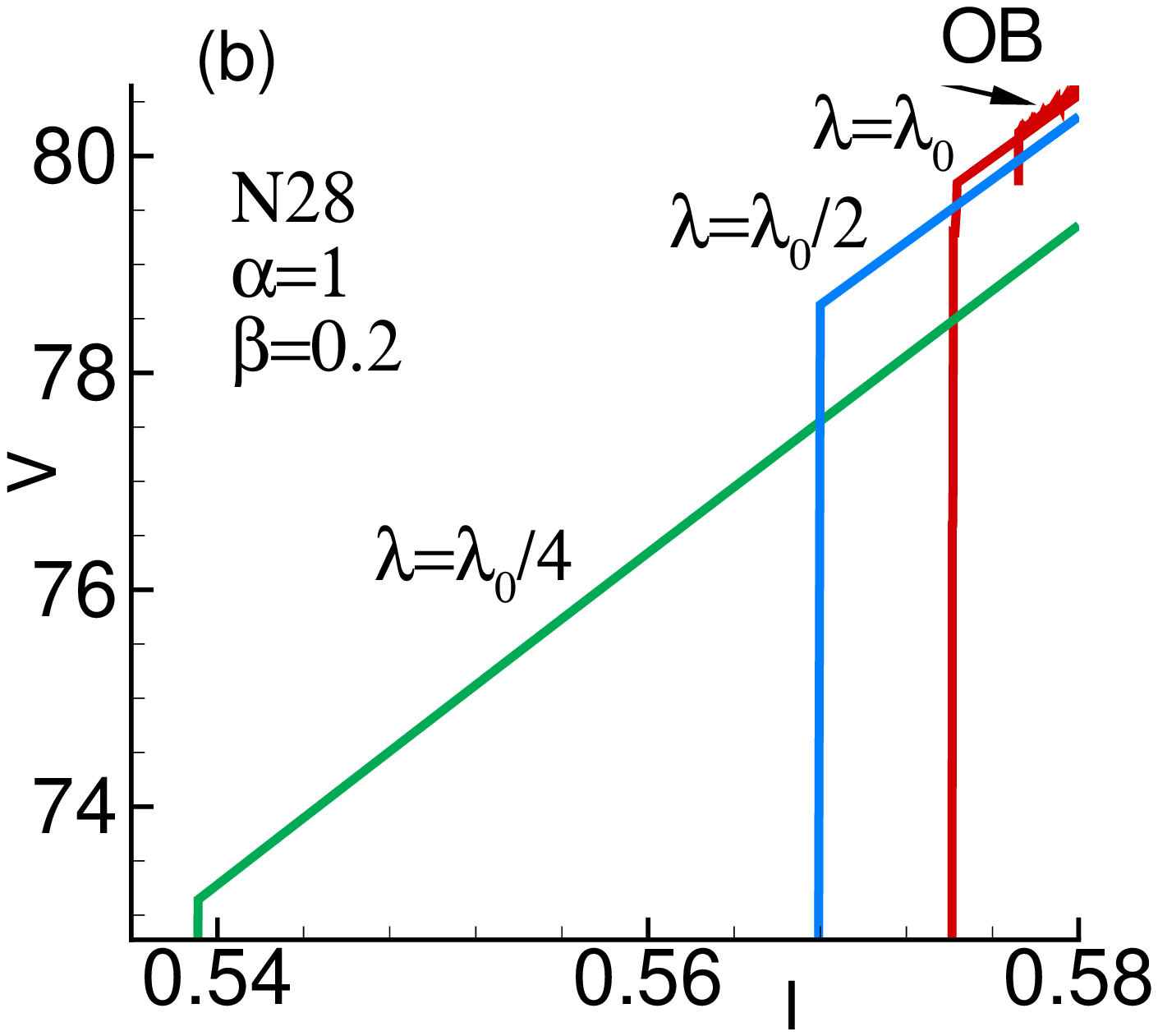}
\includegraphics[width=70mm]{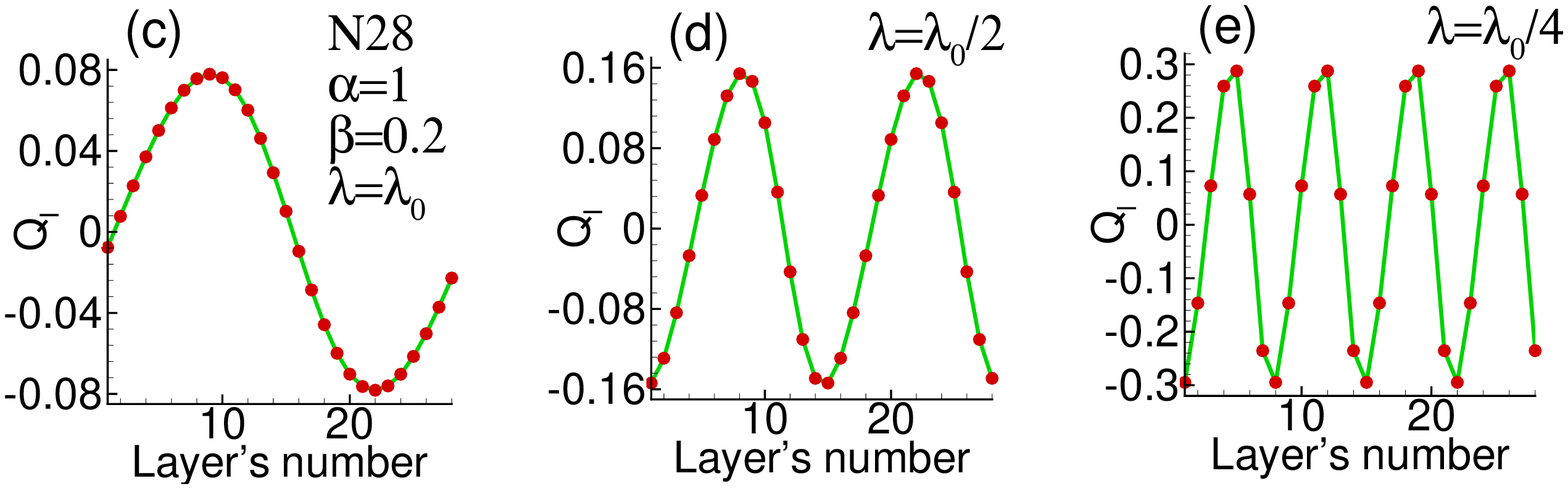}
\includegraphics[width=70mm]{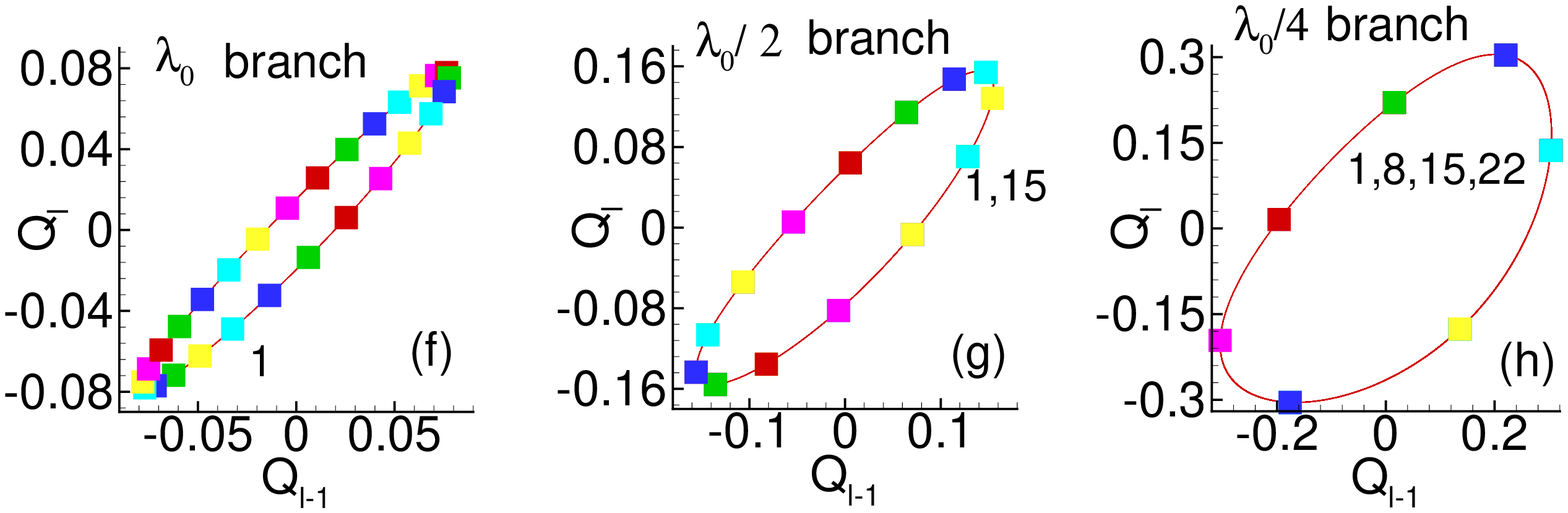}
\caption{(Color online) Demonstration of the TWB corresponding to the traveling waves with different wavelength.
(a) The manifestation of the TWB for the stack with 14 JJ. Inset demonstrates the transition from the OB to the
first TWB with $\lambda=14D$. (b) The same for the stack with 28 JJ. (c) The distribution of the charge in
S-layers along the stack with 28 JJ for TWB with $\lambda=\lambda_0$ at bias current $I=0.577$. (d) The same for
$\lambda=\lambda_0/2$. (e) The same for $\lambda=\lambda_0/4$. Figures (f)-(h) show the corresponding Lissajous
charge-charge diagrams.} \label{2}
\end{figure}

\begin{figure}
\centering\includegraphics[height=70mm]{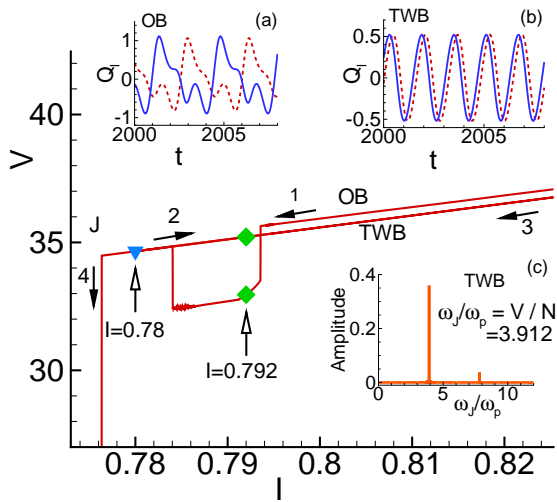} \caption{(Color online) Manifestation of the TWB in CVC of the
stack with 9 JJ at $\alpha=3$, $\beta=0.2$ and periodic BC. The insets (a) and (b) show the charge oscillations
in two S-layers at I=0.792 in OB and TWB, relatively. The inset (c) presents the results of FFT analysis of
charge oscillations for TWB  at I=0.792. }\label{3}\end{figure}

\begin{figure}
\centering\includegraphics[height=60mm]{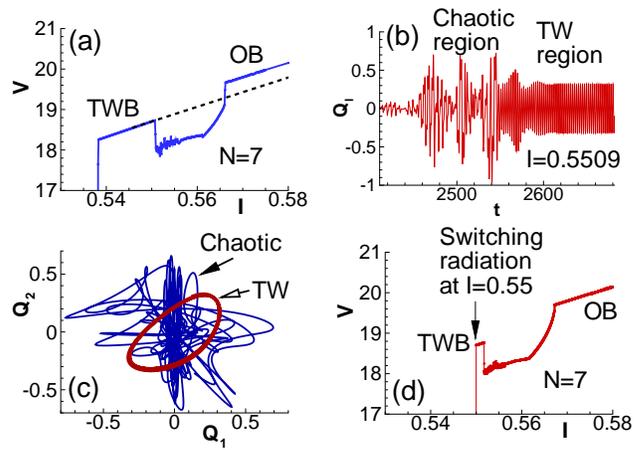} \caption{(Color online) (a) Manifestation of transition OB
$\rightarrow$ TWB in the CVC of the stack with 7 JJ. (b) The charge-time dependence  for S-layer at transition
point. (c) Charge-charge diagram for the neighbor S-layers in  chaotic region of the OB and in TWB. (d) Effect
of microwave radiation on the TWB.} \label{4}\end{figure}

The  TWB have been observed in the stacks with different number of junctions.  Particularly, the TWB in CVC were
found in the stacks with 14 JJ and 28 JJ (Fig.~\ref{2}), 9 JJ (Fig.~\ref{3}d) and 7 JJ (Fig.~\ref{4}). In case
of even number of junctions in the stack, the jump from the OB to the TWB happens first to some unstable state.
Such transition for stacks with 14 JJ at $\alpha=1$, $\beta=0.2$ is demonstrated in the inset (a) of
Fig.~\ref{2}, where \emph{a transition from the parametric resonance region of OB} is observed. For the stacks
with odd number of junctions we observe usually the jumps to the TWB from the chaotic part of breakpoint region
in OB. The increase of bias current after jump instead of its decrease allows us restore the total TWB (see
Fig.~\ref{3}).

We note that the longest wavelength of a TW in the system with N JJ is $\lambda_0=ND$, where D is a period of
lattice ( the sum of thicknesses of insulating and  superconducting layers). So, the waves with $\lambda=ND/n$
might appear in the stack, where n is devisor of number N. We stress that the TWB in Fig.~\ref{1} \emph{was
naturally obtained just by sweeping of bias current.} Another possibility is to use \emph{the variation of the
initial conditions for phase differences.} In Fig.~\ref{2} we show such realization of the TWB corresponding to
TW with different wavelengths. Particularly, in Fig.~\ref{2}a we demonstrate the manifestation of the TWB in CVC
of the stacks with 14 JJ related to the waves with $\lambda_0=14D$ and $\lambda=\lambda_0/2$. Inset demonstrates
a transition of the OB to the first TWB with $\lambda=\lambda_0$. The analysis of the time dependence of the
charge in S-layers and its distribution along the stack around the second transition in Fig.~\ref{2}a shows that
the TW with $\lambda=\lambda_0$ transforms to TW with $\lambda=\lambda_0/2$.

In Fig.~\ref{2}(b) we demonstrate the TWB  for the stack with N=28, corresponding to the states with
wavelengthes $\lambda_0=28D$, $\lambda_0/2$ and $\lambda_0/4$ . The different initial conditions for the phases
and the voltages in the JJ were used to obtain these branches. We find the wavelength of the TW  by the
distribution of the charge in the S-layers along the stack at fixed time moment. Such distribution at bias
current $I=0.577$ is shown in Fig.~\ref{2}(c),(d) and (e). These figures demonstrate also that the amplitude of
charge is greater for the TW with smaller  wavelengths.

To test if these distributions follow precisely to the TW, we present in Fig.~\ref{2}(f),(g) and (h) the
corresponding Lissajous charge-charge diagrams. In fact, two different  plots are shown in each figure: squares
show the $(Q_{l},Q_{l-1})$ for different values of $l$ at fixed time moment; the curves show the trajectory for
fixed value of $l$ in  during a whole time domain (i.e., the time domain  at constant value of current). We got
the same curve for different $l$, which means that  the charge in each layer has the  same frequency and
amplitude, but the charge in layers differers by a phase shift only. So, this wave is not a standing wave, but a
traveling one.

An example of the charge dynamics corresponding to the OB and TWB for the stack with 9 JJ at $\alpha=3$ and
$\beta=0.2$ and its simulation procedure is demonstrated in Fig.~\ref{3}. First we follow the OB (arrow 1),
decreasing the current from the value I=1.2. At I=0.78405 a transition from the chaotic part of OB to TWB is
happened. We continue the current decreasing process along TWB till the I=0.78. At this point we increase
current (arrow 2), follow the TWB till $I=5$. Then we return back (arrow 3) and follow along the TWB till the
jump to the another branch (arrow 4). The insets (a,b) show a time dependence of charge in first (solid line)
and second (dashed line) S-layers of the stack at $I=0.792$ for the OB and TWB, correspondingly. We see that in
the case of TWB the charge oscillations in  S-layers  are just shifted in phase. In the supplement we present
the animation of the charge-time dependence demonstrating the traveling wave along the stack. The inset (c)
presents the result of FFT analysis of charge-time dependence corresponding to the TWB. The time domain was
taken at I=0.792, where the value of voltage is equal to V=32.956.  As we see the value $V/N$ coincide with the
results of FFT analysis $\frac{\omega_J}{\omega_p}=3.912$.

One more manifestation of the TWB is shown in Fig.~\ref{4}, where we demonstrate a transition to the TWB for the
stack with 7 JJ at $\alpha=1,\beta=0.2$. The charge-time dependence  demonstrates a changing of character of the
charge oscillations in S-layer near a transition point at I=0.5509 (Fig.~\ref{4}b). We can see the
transformation of the chaotic behavior in the OB to the regular one in TW state. In Fig.~\ref{4}c we show the
Lissajous charge-charge diagram in chaotic (thin line) and TW (thick line)regions. The diagram demonstrate the
variation in time of the charges in two neighbor layers. The charge in the first S-layer $Q_1$ is plotted along
the x-axis and charge in the second one $Q_2$ along the y-axis. As we can see, the Lissajous charge-charge
diagram presents an open trajectory for the chaotic region and closed one for the TW region.

We find the effect of microwave radiation on TWB, which was taken into account by the including an additional
current $I_R=A*\sin{\omega_R t}$  through the stack. \emph{Increasing power of radiation (amplitude A) leads to
the destroying of the TW} and transition from the TWB to the another branch in CVC. In Fig.~\ref{4}d we
demonstrate the effect of radiation, which was switch on  at $I=0.55$ with amplitude $A=1$ and Josephson
frequency $\omega_J$. As we can see, the TWB disappears under this radiation. This result open a way for the
experimental testing the observed branches in CVC.
\begin{figure} \centering\includegraphics[height=65mm]{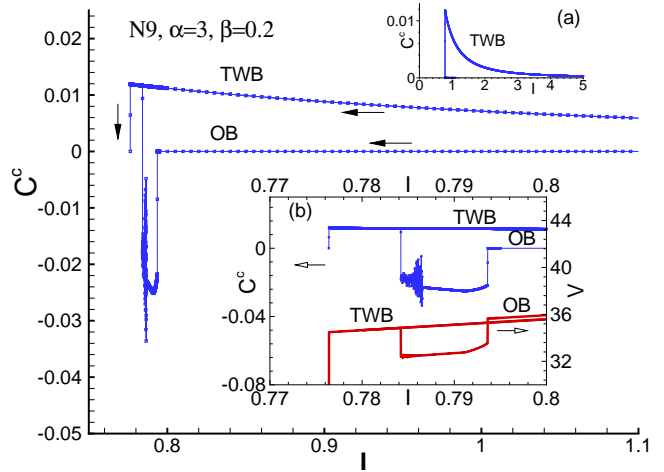}\caption{(Color online) The dependence of correlation function
"charge-charge" $C^c$ for neighbor layers three and four on the bias current. The procedure of simulation is the
same as for Fig.~\ref{3}. Inset (a) stress the growth of $C^c$ for the TWB in the bigger interval of I. Inset
(b) enlarged the $C^c(I)$ dependence near the transition from the OB to the TWB. The thick curve show the CVC in
this transition region.}\label{5}\end{figure}

To compare dynamical behavior of the states corresponding to the OB and TWB in long bias current interval, we
have made a correlation functions analysis\cite{shk-prb09}. We have investigated a correlation of charge in
neighboring layers by charge-charge correlation functions
$C_{l,l+1}^c=<Q_l(t)Q_{l+1}(t)>=lim_{Tm-Ti\rightarrow\infty}\frac{1}{Tm-Ti}\int_{Ti}^{Tm} {Q_l(t)Q_{l+1}(t)}dt$.
Fig.~\ref{5} shows the dependence of correlation function "charge-charge" $C^c$ for neighbor layers three and
four $C^c_{4,3}$ on bias current  for the OB and the TWB. We stress that \emph{all curves were obtained in the
\emph{one} numerical experiment.} Inset (a) demonstrate  the increase of correlations ($C^c$) in the TW state
with decreasing of the bias current in the large interval of $I$. As it has been seen from the inset (b) which
enlarges the region near BP, the correlation functions $C^c$ for the OB demonstrate the characteristic BPR
structure with the features obtained in the Ref.\cite{shk-prb09}. We see also that the features of correlation
functions coincide with the features of CVC.  The $C^c_{l,l-1}$ for TWB are coincide for all $l$. The
correlation functions current-current (not presented here) show the same behavior. The autocorrelation functions
for TW state, which are  defined by
$C_l^a=<Q_l(t)Q_l(t-t_1)>=lim_{Tm-Ti\rightarrow\infty}\frac{1}{Tm-Ti}\int_{Ti}^{Tm}Q_l(t)Q_l(t-t_l)dt$, have a
constant amplitude, i.e., the system in the state with TW is complectly self repeating.
\begin{figure} \centering\includegraphics[height=65mm]{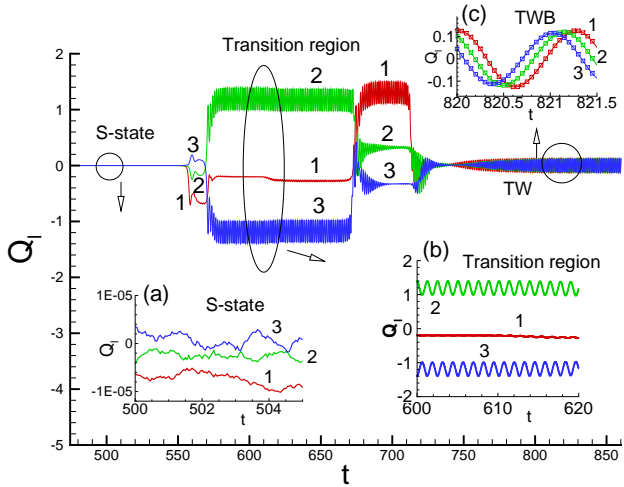}\caption{(Color online) The charge-time dependence
in transition region near $I=I_c$ (see the inset (b) in Fig.~\ref{1}). The inset (a) enlarges the $Q_l(t)$
dependence in three first S-layers in zero-voltage state. The inset (b) demonstrates the $Q_l(t)$ dependence for
the same S-layers in the transition region shown by oval. The inset (c) shows the  $Q_l(t)$ dependence for the
same layers in the state with TW.}\label{6}\end{figure}

Finally, let us describe shortly the creation of the TWB at  $I=I_c$. In Fig.~\ref{6} we present the charge-time
dependence in transition region around $I=I_c$ (see the inset (b) in Fig.~\ref{1}) for three S-layers in the
stack with 10 JJ.  We enlarged the $Q_l(t)$ dependence in this region and show it in three insets separately:
the inset (a) shows it in zero-voltage state, inset (b) - in some part of the transition region shown by oval,
inset (c) - in the state with TW. The simulation was done by increase a bias current in the interval
$(0.993,1.003)$ around $I=I_c$ with a step in current $\delta I = 10^{-5}$. We observe here a complex dynamical
behavior in a very short time interval inside of the time domain. Fluctuations of charge in S-layers in zero
voltage state is replaced by specific changes of the charge and  the oscillations with nonzero time averaged
value (charge value in S-layer is high enough, up to $|Q|=2.5$), then we see a transition to the TW state. The
amplitude of charge oscillations in TW state is large, but its average in time is zero. A detailed description
of the switching dynamics will be presented somewhere else.

In summary, we showed that CVC  of coupled Josephson junctions has the branches related to the charge traveling
waves along the c-axis. Transitions between such branches demonstrating the changing of wavelength of the
traveling wave with a decrease of bias current  were found. We fixed also the transitions from the outermost
branch to the traveling wave branch. Detailed analysis of the time dependence of the charge in superconducting
layer, its FFT analysis and the investigations of correlation functions showed different features of such
transitions, particularly, different behavior for the outermost and traveling wave branches. These results shed
light on the resonance features of the coupled Josephson junctions which are investigated intensively today. The
detailed analysis of the branch structure in the experimental CVC of IJJ has not been done yet. We proposed a
method to distinguish the branch related to the charge traveling wave along c-axis from the other branches in
CVC by microwave irradiation.

We thank R. Kleiner, A. Yurgens, V. Krasnov, M. Suzuki,  F. Mahfouzi and M. R. Kolahchi  for helpful
discussions.

\end{document}